\documentstyle[aps,epsfig,prl]{revtex}

\begin{document}

\draft

\title{Vibrational spectrum of topologically disordered systems}

\author{T.~S.~Grigera$^1$, V.~Mart\'{\i}n-Mayor$^1$, G.~Parisi$^1$ and
P.~Verrocchio$^2$}

\address{$^1$ Dipartimento di Fisica, Universit\`a di Roma ``La
Sapienza'', Piazzale Aldo Moro 2, 00185 Roma, Italy \\ INFN sezione di
Roma - INFM unit\`a di Roma \\ $^2$ Dipartimento di Fisica, Universit\`a
di Trento, Via Sommarive, 14, 38050 Povo, Trento, Italy \\ INFM unit\`a
di Trento}

\date{May 18, 2001}

\wideabs{
\maketitle

\begin{abstract}
The topological nature of the disorder of glasses and supercooled
liquids strongly affects their high-frequency dynamics.  In order to
understand its main features, we analytically studied a simple
topologically disordered model, where the particles oscillate around
randomly distributed centers, interacting through a generic pair
potential.  We present results of a resummation of the perturbative
expansion in the inverse particle density for the dynamic structure
factor and density of states. This gives accurate results for the
range of densities found in real systems.
\end{abstract}

}

The high frequency dynamics of glasses and supercooled liquids has
recently received a large amount of
experimental~\cite{ESPERIMENTI,secondary-peak,marchmeeting},
numerical~\cite{SIMULAZIONI,Ruo2,HoKoBi}, and
theoretical~\cite{GoMa,MePaZe,MaMePaVe,TARASKIN,lattice,MaPaVe}
attention.  High-resolution inelastic X-ray scattering techniques have
made accessible to experiment the region where the exchanged external
momentum $p$ is comparable to $p_0$, the maximum of the static
structure factor.  A number of facts have emerged from these
experiments: a) The dynamic structure factor (DSF) $S(p,\omega)$ has a
Brillouin-like peak for momenta up to $p/p_0 \sim 0.1-0.5$, usually
interpreted in terms of propagating acoustic-like excitations, whose
velocity extrapolates to the macroscopic sound velocity when $p \to
0$.  b) The peak has a width $\Gamma$, due to disorder, which in a
large variety of materials seems to scale as $\Gamma = A p^{\alpha}$,
with $\alpha \sim 2$ and $A$ depending very slightly on the
temperature.  c) The density of states (DOS) exhibits an excess
respect to the Debye behavior ($g(\omega)\propto\omega^2$), known as
the {\em Boson peak}, which is most remarkable for strong glasses.

From the theoretical viewpoint, the challenge is to explain the above
features of the scattering spectra.  For a dense system like a glass,
the short time dynamics is naturally interpreted as a consequence of
vibrations around a quenched disordered structure.  Indeed, molecular
dynamics simulations have shown that {\em harmonic vibrations} are
enough to describe the dynamic structure factor (DSF)~\cite{Ruo2} and
specific heat~\cite{HoKoBi}.  Nevertheless, the analytical problem is
very hard even in this first approximation, because it involves a
matrix (the Hessian) with random elements.  So the study of vibrations
in glasses is related to the general problem of analyzing the
statistical properties of large random matrices~\cite{Metha}.  This is
a venerable problem, with applications that include the theory of
nuclear spectra, conductivity in alloys, and many others~\cite{Metha}.
We should distinguish between matrices obtained as a result of random
perturbations of a reference lattice, and those for which no reference
lattice can be defined~\cite{Zim}.  Lattice-based random matrices
arise typically in problems of the solid state (Anderson localization,
transport in mesoscopic systems, etc.).  A number of well-known
approximations have been developed to deal with these kind of
matrices: this approximations can be derived by considering the
disordered part of the random matrix as a perturbation of the solvable
crystalline problem~\cite{Zim} and partially resumming the
perturbative expansion.  On the other hand, {\em off-lattice} random
matrices arise in many problems, such as amorphous semiconductors,
instantaneous normal modes in liquids, and the vibrational excitations
of glasses.  These matrices lack a natural separation between a
solvable ordered part and a disordered perturbation.  As a
consequence, the theoretical tools for their study have started to be
developed much more recently~\cite{Bassa}, and are best suited for low
densities, with emphasis on the DOS rather than the DSF. Furthermore,
the Debye behavior of the DOS is not reproduced by the above methods,
even in the glassy phase (see Cavagna et al.\ in~\cite{Bassa}).

The purpose of this Letter is to address the problem of harmonic
vibrations in glasses and to find a self-consistent integral equation.
This equation is somewhat analogous to the lattice-based
approximations mentioned above, but is appropriate for the off-lattice
matrices that arise in this case~\cite{MePaZe}. Our approach is based
on an expansion in $1/\rho$ ($\rho$ is the number of particles per
unit volume)~\cite{MePaZe,MaMePaVe}.

Somehow complementary to the present approach is a recent modification
of Mode Coupling Theory~\cite{GoMa} to describe excitations around a
quenched structure, which has been applied to a hard-sphere fluid.
Also, lattice-based models have been proposed~\cite{TARASKIN,lattice},
but it turns out~\cite{MaPaVe} that they definitely miss the $p^2$
behavior of the peak width. Other differences between the lattice and
off-lattice cases regarding the DOS will be discussed.


We consider a model of particles oscillating harmonically around
(disordered) equilibrium positions.  For simplicity we assume that
they can move only along a direction ${\mbox{\boldmath$u$}}$: \hbox{
${\mbox{\boldmath$x$}}_j(t)={\mbox{\boldmath$x$}}_j^{\mathrm{eq}}+
{\mbox{\boldmath$u$}}\varphi_j(t)$}, $j=1,\ldots,N$.  The vibrational
potential energy is then $ V(\{\varphi_i\})=\frac{1 }{2}\sum_{i,j}
f({\mbox{\boldmath$x$}}_i^{\mathrm{eq}} -
{\mbox{\boldmath$x$}}_j^{\mathrm{eq}}) (\varphi_i - \varphi_j)^2$,
where $f$ is the second derivative of the pair potential.  Of interest
is the spectrum of the dynamical matrix
(Hessian)~\cite{MaMePaVe,Proceedings}
\begin{equation}
M_{ij}=\delta_{ij} \sum_{k=1}^N
f({\mbox{\boldmath$x$}}_i^{\mathrm{eq}}-{\mbox{\boldmath$x$}}_k^{\mathrm{eq}})
-
f({\mbox{\boldmath$x$}}_i^{\mathrm{eq}}-{\mbox{\boldmath$x$}}_j^{\mathrm{eq}}).
\label{HESSIANO}
\end{equation}

Neglecting quantum effects, in the harmonic approximation the DSF can
be obtained as (see e.g.~\cite{MaMePaVe})
\begin{equation}
S(p,\omega)= {k_{\mathrm B}T p^2 \over m \omega^2} \overline{\sum_n
 \left| \sum_i e_{n,i} e^{i \mbox{\boldmath\scriptsize$p$} \cdot
 {\mbox{\boldmath\scriptsize$x$}}^{\mathrm eq}_i}
 \right|^2
 \delta(\omega-\omega_n) } ,
\end{equation}
where $e_{n,i}$ is the $i$-th component of the $n$-th eigenvector of
the Hessian (\ref{HESSIANO}), the eigenfrequencies $\omega_n$ are the
square root of the corresponding eigenvalues, $T$ is the temperature
and the overline stands for the average over
$\{{\mbox{\boldmath$x$}}_j^{\mathrm{eq}}\}$ (hereafter we set the mass
$m=1$).

The model is defined by the probability distribution of the
$\{{\mbox{\boldmath$x$}}_j^{\mathrm{eq}}\}$ and by the function
$f(r)$.  Obviously, in a real system the
$\{{\mbox{\boldmath$x$}}_j^{\mathrm{eq}}\}$ are highly correlated.
However, as discussed in detail in~\cite{MaMePaVe,Proceedings}, it is
a good approximation to take
$\{{\mbox{\boldmath$x$}}_j^{\mathrm{eq}}\}$ uncorrelated while at the
same time using an effective interaction which weights the physical
$f$ with the radial distribution function $g(r)$.  For a given $f$,
the only free parameter is $\rho$.

As usual, we consider the resolvent $ G(p,z) \equiv \sum_{jk}
\overline{ \exp[i \mbox{\boldmath\scriptsize$p$} \cdot (
{\mbox{\boldmath\scriptsize$x$}}^{\mathrm eq}_j-
{\mbox{\boldmath\scriptsize$x$}}^{\mathrm eq}_k ) ] [z-M]^{-1}_{jk} }
$, since the DSF can be obtained as $S(p,\omega)= - 2 k_{\mathrm B} T
p^2 {\mathrm Im}\, G(p,\omega^2 + {\mathrm i} 0^+) / (\omega \pi)$.
Our aim is to compute $G$ using the appropriate self-consistent
equations.  Although here there is no separation between an ordered
reference state and a random perturbation, it has been
observed~\cite{MePaZe} that in the limit of infinite density the
spectrum coincides with that of an elastic homogeneous medium, since
the resolvent is simply $G(p,z) = [z-\lambda(p)]^{-1}$ (where
$\lambda(p) = \rho[\hat f(0)-\hat f(p)]$ and $\hat f(p)$ is the
Fourier transform of $f(r)$).  A systematic perturbative expansion in
$1/\rho$ has been constructed in ref.~\cite{MaMePaVe}.  Technically,
an expansion is found for the resolvent, which is written as
\begin{equation}
G(p,z) =  \frac{1}{z-\lambda(p)-\Sigma(p,z)}\,.\label{GDIPI}
\end{equation}
The self-energy $\Sigma(p,z)=\Sigma'(p,z) + {\mathrm i} \Sigma''(p,z)$
is then expanded in powers of $1/\rho$ \cite{MePaZe,MaMePaVe} in the
relevant region where $\rho=O(z)$. At infinite density, $\Sigma=0$ and
the dispersion relation is $\omega(p)=\sqrt{\lambda(p)}$, which is
linear at small $p$.  The real self-energy $\Sigma'(p,z)$ renormalizes
the dispersion relation, while the imaginary part gives the peak
width: $\Gamma = \Sigma''(p,\omega^2_p)/\omega_p$, where $\omega_p$ is
the position of the maximum of the DSF.

\begin{figure}
\epsfig{file=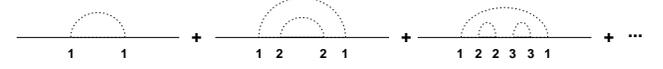, angle=0, width=\columnwidth}
\caption{The diagrams of the $1/\rho$ expansion which are taken into
account in our approach. The numbers correspond to the particle-label
repetitions}
\label{DIAGRAMMI}
\end{figure}

We have reformulated the $1/\rho$ expansion~\cite{INPREPARATION} in a
way that reduces the number of diagrams and allows to identify those
with the simple topology of Fig.~\ref{DIAGRAMMI}. Topologically, these
diagrams are exactly those considered in the usual lattice CPA and in
other self consistent approximations.  The sum of this infinite subset
is given by the solution of the integral equation:
\begin{equation}
\Sigma(p,z)=\frac{1}{\rho}\int \!\! \frac{d^3 q}{(2\pi)^3} 
\left[\rho\left(\hat f(\mbox{\boldmath$q$})-\hat f(\mbox{\boldmath$p$}-
\mbox{\boldmath$q$})\right)\right]^2 G(p,z),
\label{CACTUS}
\end{equation}
where the resolvent is given by Eq.~(\ref{GDIPI}). The solution gives
us the resolvent, and hence the DSF and DOS (Eq.~\ref{PGRANDE} below).

We are interested in the solution of Eq.~(\ref{CACTUS}) for different
values of $z$ and $\rho$.  To be definite, we consider an explicit
case where the function $f(r)$ has a simple form, namely
$f(r)=\exp[-r^2/(2\sigma^2)]$.  This is a reasonable first
approximation for the effective interaction~\cite{Proceedings}.  We
shall take $\sigma$ as the unit of length and set $p_0=1/\sigma$,
which is a reasonable choice for $p_0$ for this Gaussian $f(r)$, as
discussed in~\cite{Proceedings}.  In this particular case we will
solve numerically the self-consistence equation.  We will also
evaluate by simulation (using the method of moments~\cite{MOMENTI})
the exact DSF and DOS by computing the resolvent for concrete
realizations of the dynamical matrix, considering a sufficiently high
number of particles so that finite volume effects can be neglected.
These numerical results will be supplemented by analytic results, that
are $f$-independent and can be obtained in the limits $p\to\infty$ and
$p\to 0$.

The infinite momentum limit is particularly interesting because of the
remarkable result~\cite{MaMePaVe} that the DOS $g(\omega)$ can be
written as
\begin{equation}
    g(\omega) = \lim_{p\to\infty} \frac{ \omega^2 S^{(1)}(p,\omega) }
{k_{\mathrm B} T p^2}.
\label{PGRANDE}
\end{equation}
We easily find that in this limit Eq.~(\ref{CACTUS}) can be written as
(${\cal G}(z)=\lim_{p\to\infty} G(p,z)$ and ${\cal A}= (2\pi)^{-3}\int
\!\!  \hat f^2(\mbox{\boldmath$q$})\,d^3q$):
\begin{equation}
\frac{1}{\rho{\cal G}(z)}=\frac{z}{\rho}-\hat f(0) -{\cal A} {\cal G}(z) -\int
\!\!\! \frac{d^3q}{(2\pi)^3}\hat f^2(\mbox{\boldmath$q$})
G(\mbox{\boldmath$q$},z)
\label{DENSIDADDEESTADOS}
\end{equation} 
A simple approximation consists in neglecting the last term in the
r.h.s. of (\ref{DENSIDADDEESTADOS}), which is reasonable at large $z$.
This approximation implies a DOS which is semicircular as a function
of $\omega^2$, with width proportional to $\sqrt{\rho}$ and centered
at $\omega^2=\rho\hat f(0)$. Translational invariance also requires
low-frequency modes. These are given by the neglected term, and in
fact it is easy to show that at high density it produces a Debye
spectrum which extends between zero frequency and the semicircular
part.

In the limit $p\to0$, the leading contribution to $\Sigma''$ comes
from $q \gg p$ in Eq.~(\ref{CACTUS}), where $G(q,z) \approx {\cal
G}(z)$, so we can write for the peak width $\Gamma(p) \approx
\Gamma_0(p)$, where
\begin{equation}
\Gamma_0(p) \equiv
\pi \rho \frac{g(\omega_p)}{2\omega_p^2}
\int \!\!\! \frac{d^3 q}{(2\pi)^3}
\left[\hat f(\mbox{\boldmath$q$})-\hat f(\mbox{\boldmath$p$}-
\mbox{\boldmath$q$})\right]^2.
\label{sigma-infinito}
\end{equation}
The integral is of order $p^2$, so if the spectrum is Debye-like for
small frequencies, we get $\Gamma(p) \sim p^2$.

\begin{figure}
\epsfig{file=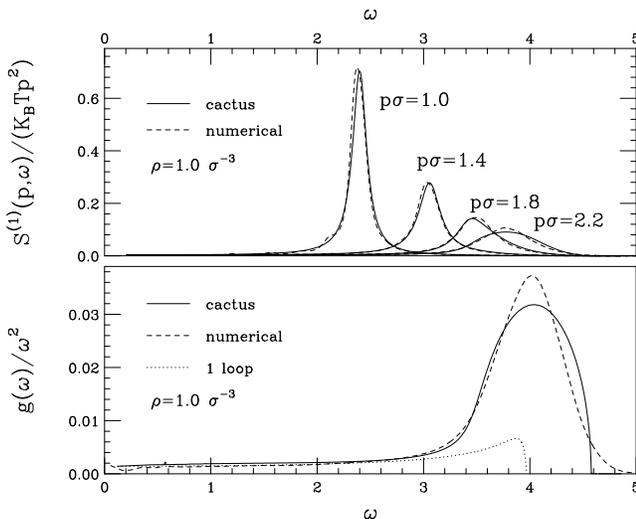, angle=90, width=\columnwidth}
\caption{Top: dynamic structure factor as obtained from
Eq.(\protect\ref{CACTUS}) (full line) and from
simulations~\protect\cite{MaMePaVe} (dashes). Bottom: DOS divided by
$\omega^2$ (Debye behavior) as obtained from
Eq.(\protect\ref{DENSIDADDEESTADOS}) (full line), simulations
(dashes), and first order in the $1/\rho$ expansion (dots).}
\label{RHO1}
\end{figure}

These considerations are verified by the numerical solution of the
Gaussian case, which are shown in Fig.~\ref{RHO1} for $\rho=1.0\,
\sigma^{-3}$ together with the results for the
simulations~\cite{MaMePaVe}. Note the good agreement, to be expected
for high-densities, and how, for large $p$, $S^{(1)}(p,\omega)$
(Fig.~\ref{RHO1}, top) tends to the DOS.  The DOS from the
self-consistent equation (Fig.~\ref{RHO1},bottom) also agrees very
well with the results from simulations, and is a big improvement over
the first term of the expansion in powers of $\rho^{-1}$. The two
contributions (Debye and semicircle) mentioned above can be clearly
identified. Our approximation fails in reproducing the exponential
decay of the DOS at high frequencies, which is non perturbative in
$1/\rho$~\cite{Zee2} and corresponds to localized states.

\begin{figure}
\epsfig{file=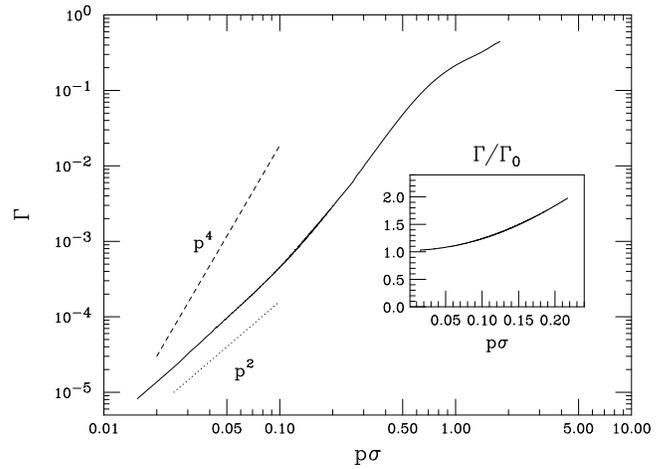, angle=90, width=\columnwidth}
\caption{Peak width vs.\ $p$, for $\rho=0.6\sigma^{-3}$. The inset
shows that $\Gamma_0(p)$ is the dominant contribution at small $p$.}
\label{scaling}
\end{figure}

Next in Fig.~\ref{scaling} we plot the linewidth as a function of $p$
as obtained from Eq.(\ref{CACTUS}). Notice that we recover the
behavior predicted from the first two non-trivial terms in the
expansion in powers of $\rho^{-1}$~\cite{MaMePaVe}: the linewidth is
proportional to $p^2$ at small $p$ (also predicted by the argument
above), then there is a faster growth and finally it approaches to a
constant as $S^{(1)}(q,\omega)$ starts to collapse onto the DOS. The
inset shows that the contribution Eq.~(\ref{sigma-infinito}) is indeed
dominant at small $p$.  However, rigorous $p^2$ scaling is found only
for very small momenta ($p/p_0<0.1$), while experiments are done at
$0.1<p/p_0<1$. In this crossover region, our model predicts deviations
from $p^2$, which are probably hard to measure experimentally. In any
case, the effective exponent is certainly less than 4, in contrast
with lattice models and consistent with experimental findings. Similar
conclusions can be drawn from mode coupling theory (Fig.~8 of
\cite{GoMa}).

\begin{figure}
\epsfig{file=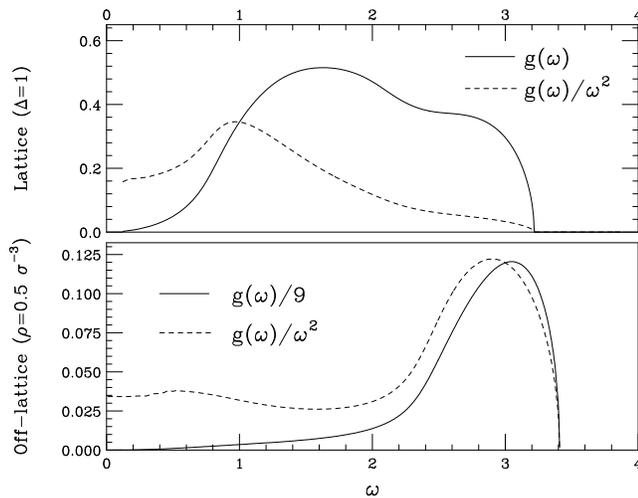, angle=90, width=\columnwidth}
\caption{Top: $g(\omega)$ (full line) and $g(\omega)/\omega^2$
(dashes) of the model of Ref.~\protect\cite{TARASKIN}, as predicted by
the single-link CPA, at $\Delta=1$. Bottom: $g(\omega)/9$ (full line)
and $g(\omega)/\omega^2$ (dashes) for our model at $\rho=0.5
\sigma^{-3}$. Note that for the values of the parameters shown here,
the lattice model is nearer to its critical point
($(\Delta_c-\Delta)/\Delta_c \approx 0.22$ while $(\rho-\rho_c)/\rho_c
\approx 0.61$).}
\label{inestabilidad}
\end{figure}

Finally, let us discuss the low density case. As mentioned,
Eq.~(\ref{DENSIDADDEESTADOS}) tells us that the position of the center
of the semicircular peak in the DOS goes to 0 as $\omega^2\sim\rho$,
but with a width that only decreases as $\sqrt{\rho}$. So at low
enough densities the DOS develops a tail that extends into negative
values of $\omega^2$. In our Gaussian case, this starts happening at
$\rho_{\mathrm{c}}=0.31\sigma^{-3}$. Since our original Hessian is
positive definite, one could regard such an instability as just a
failure of the $1/\rho$ expansion. However, the structure that
Eqs.~(\ref{CACTUS}) and~(\ref{DENSIDADDEESTADOS}) (wrongly) predict
from the hybridization between the Debye and semicircle spectra for
$\rho\gtrsim\rho_{\mathrm{c}}$ is extremely suggestive of a
Boson-Peak. In fact, it can be shown~\cite{INPREPARATION} that for
$\rho\gtrsim\rho_{\mathrm{c}}$, and very small values of $\omega$ a
cross-over occurs from the Debye $\omega^2$ behaviour at low $\omega$
to a $\omega^{3/2}$. This structure (Fig.~\ref{inestabilidad}, bottom)
can be compared with the Boson peak predicted by the lattice
integral-equations~\cite{TARASKIN,lattice}, close to their
instabilities (see Fig.~\ref{inestabilidad}, top).  The Boson peak
predicted by the off-lattice equations is a subtle feature of the
spectrum that can only be seen when dividing the DOS by the Debye
behavior. It is also striking that this off-lattice Boson peak shifts
towards $\omega=0$ when the instability is approached, as the
experimental peak does upon heating. On the other hand, it is fair to
say that the Boson peak of the lattice integral equations is just the
first maximum of the DOS (which experimentally is not very
temperature-dependent), shifted to low $\omega$ by the division by
$\omega^2$.

In conclusion, we have presented a random-matrix approach to study an
off-lattice model for the vibrational dynamics of glasses, in which
particles oscillate harmonically along a given direction around
quenched equilibrium positions.  At variance with disordered systems
defined as a perturbation of a reference crystal, our system lacks a
natural separation between a {\em solvable} and a {\em
random-perturbation} term. Nevertheless, an expansion in $1/\rho$ can
be obtained~\cite{MaMePaVe}. Here we have reformulated this expansion
in a way that allows us to resum an infinite subset of terms,
obtaining an integral self-consistent equation (\ref{CACTUS}) for the
dynamic structure factor and the DOS. This equation predicts a $p^2$
scaling of the linewidth for a generic potential, as well as the
appearance of a Boson-peak-like structure at low densities, linked to
a mechanical instability.  Since the dynamic structure factor tends to
the DOS at large momentum, we argue that Boson-peak-like features
found experimentally in the former at large
$p$~\cite{secondary-peak,marchmeeting} are just another manifestation
of the Boson peak.  We compared the results of eq.~(\ref{CACTUS}) with
numerical calculations for a gaussian potential and found satisfactory
agreement at densities comparable to those of real glasses. At low
densities, the resummation finds a spectrum at negative squared
frequencies, which is unphysical for this potential. However, for
$\rho\gtrsim\rho_{\mathrm{c}}$ a structure appears which is definitely
different from those found in lattice models~\cite{TARASKIN,lattice},
and extremely remindful of the experimental Boson peak.

These results can be extended in several ways (potentials with
instabilities at low densities, non-collinear vibrations, detailed
study of the correlated case); work in these directions is in
progress.

We thank J.\ L.\ Alonso, L.\ A.\ Fern\'andez and S.\ A.\ Grigera for
critical reading of the manuscript.  We acknowledge partial support
from CICyT, M.E.C. (Spain) (VMM), European Commission (VMM) and CONICET
(Argentina) (TSG).


\begin{thebibliography}{99}

\bibitem{ESPERIMENTI} P.\ Benassi et al., Phys.\ Rev.\ Lett.\ {\bf
77,} 3835 (1996); C.\ Masciovecchio et al., Phys.\ Rev.\ B {\bf 55,}
8049 (1997); C.\ Masciovecchio et al., Phys.\ Rev.\ Lett., {\bf 76,}
3356 (1996); G.\ Monaco, et al., Phys.\ Rev.\ Lett.\ {\bf 80,} 2161
(1998); C.\ Masciovecchio et al., Phys.\ Rev.\ Lett.\ {\bf 80,} 544
(1998); D.\ Fioretto et al., Phys.\ Rev.\ E {\bf 59,} 4470 (1999); P.\
A.\ Sokolov et al., Phys.\ Rev.\ E {\bf 60,} 2464 (1999); G.\ Ruocco
et al., Phys.\ Rev.\ Lett.\ {\bf 83,} 5583 (1999); F.\ Sette et al.,
Science {\bf 280,} 1550 (1998); U.\ Buchenau et al., Phys.\ Rev.\ B
{\bf 34,} 5665 (1986); M.\ Foret, et al., Phys.\ Rev.\ Lett.\ {\bf
77,} 3831 (1996).

\bibitem{secondary-peak} F.\ Sette et al., Phys.\ Rev.\ Lett.\ {\bf
77,} 83 (1996).

\bibitem{marchmeeting} B.\ Hehlen et al., invited talk at the {\em APS
March meeting 2001,} Seattle; L.\ Brjeson, et al., {\em ibid.}


\bibitem{SIMULAZIONI} J.\ Horbach et al., cond-mat/9910445; S.\ N.\
Taraskin and S.\ R.\ Elliot, Phys.\ Rev.\ B {\bf 59,} 8572 (1999); R.\
Dell'Anna et al., Phys.\ Rev.\ Lett.\ {\bf 80,} 1236 (1998); J.\ L.\
Feldman et al., Phys.\ Rev.\ B {\bf 59,} 3551 (1999); P.\ L.\ Allend
et al., cond-mat/9907132; M.\ C.\ C.\ Ribeiro et al., J.\ Chem.\
Phys.\ {\bf 108,} 9027 (1998); V.\ Mazzacurati et al., Europhys.\
Lett.\ {\bf 34,} 681 (1996); M.\ Sampoli, et al., Philos.\ Mag.\ B {\bf
77,} 473 (1998).


\bibitem{Ruo2} G.\ Ruocco et al., Phys.\ Rev.\ Lett. {\bf 84,} 5788 (2000).

\bibitem{HoKoBi} J.\ Horbach et al., J.\ Phys.\ Chem.\ B
{\bf 103,} 4104 (1999).


\bibitem{GoMa} W.\ G\"{o}tze and M.\ R.\ Mayr, Phys.\ Rev.\ E {\bf
61,} 587 (2000).


\bibitem{MePaZe} M. M\`ezard et al., Nucl.\ Phys.\ {\bf B559,} 689
(1999).


\bibitem{MaMePaVe} V.\ Mart\'\i{}n-Mayor et al., J.\ Chem.\ Phys.\
{\bf 114}, 8068 (2001).


\bibitem{TARASKIN} S.\ N.\ Taraskin et al., Phys.\ Rev.\ Lett.\ {\bf
86,} 1255 (2001).

\bibitem{lattice} W.\ Schirmacher et al., Phys.\ Rev.\ Lett.\ {\bf
81}, 136 (1998); J.\ W.\ Kantelhardt et al., Phys.\ Rev.\ B {\bf 63}
064302 (2001).

\bibitem{MaPaVe} V.\ Mart\'\i{}n-Mayor et al., Phys.\ Rev.\ E {\bf
62,} 2373 (2000).


\bibitem{Metha} M.\ L.\ Metha, {\em Random matrices}, Academic Press
(1991); T.\ Guhr, A.\ M\"uller-Groeling, H.\ A.\ Weidenm\"uller,
Phys.\ Rep.\ {\bf 299}, 189 (1998).


\bibitem{Zim} See {\em e.g.} J.\ M.\ Ziman, {\em Models of disorder},
Cambridge University Press, Cambdrige (1979).

\bibitem{Bassa} A.\ J.\ Bray and G.\ J.\ Rodgers, Phys. Rev. B {\bf
38}, 11461 (1988); T.\ M.\ Wun and R.\ F.\ Loring, J. Chem. Phys {\bf
97,} 8368 (1992); Y.\ Wan and R.\ Stratt, J. Chem. Phys {\bf 100,}
5123 (1994); T.\ Keyes, J.\ Phys.\ Chem.\ A {\bf 101,} 2921 (1997); A.\
Cavagna et al., Phys. Rev. Lett. {\bf 83,} 108 (1999); G.\ Biroli, R.\
Monasson, J.\ Phys. A: Math. Gen. {\bf 32,} L255 (1999).


\bibitem{Proceedings} T.\ S.\ Grigera et al., cond-mat/0104433 (to be
published in Philos. Mag. {\bf B}).

\bibitem{INPREPARATION} T.\ S.\ Grigera et al., to be published.
 
\bibitem{MOMENTI} C.\ Benoit et al., J.\ Phys.:
Condens.\ Matter {\bf 4}, 3125 (1992) and references therein; P.\
Turchi et al., J.\ Phys.\ C {\bf 15,} 2891
(1982).

\bibitem{Zee2} A.\ Zee and I.\ Affleck, J.\ Phys: Condens.\ Matter {\bf 12},
8863 (2000).

\end{thebibliography}
\end{document}